\documentclass[%
 reprint, twocolumn
 amsmath,amssymb,
 aps,
]{revtex4-2}

\usepackage{graphicx}
\usepackage{dcolumn}
\usepackage{bm}
\usepackage{hyperref}
\usepackage[mathlines]{lineno}
\usepackage{amsmath}

\begin{document}

\preprint{APS/123-QED}

\title{Low-loss, fabrication-tolerant, and highly-tunable Sagnac loop reflectors and Fabry–Pérot cavities on thin-film lithium niobate }

\author{Luke Qi$^{1,*}$}
\author{Ali Khalatpour$^{1}$}
\author{Jason Herrmann$^{1}$}
\author{Taewon Park$^{2}$}
\author{Devin Dean$^{1}$}
\author{Sam Robison$^{1}$}
\author{Alexander Hwang$^{1}$}
\author{Hubert Stokowski$^{1}$}
\author{Darwin Serkland$^{3}$}
\author{Martin Fejer$^{1}$}
\author{Amir H. Safavi-Naeini$^{1,*}$}

\affiliation{$^1$Department of Applied Physics and Ginzton Laboratory, Stanford University, Stanford, California 94305, USA}

\affiliation{$^{2}$Department of Electrical Engineering and Ginzton Laboratory, Stanford University, Stanford, California 94305, USA}

\affiliation{$^{3}$Sandia National Laboratories, Albuquerque, New Mexico 87185, USA}

\affiliation{$^{*}$lukeqi7,safavi@stanford.edu}

\date{\today}

\begin{abstract}
We present low-loss ($<1.5\%$) and power-efficient Mach-Zehnder interferometers (MZIs) on thin-film lithium niobate. To accurately measure low MZI losses, we develop a self-calibrated method using tunable Sagnac loop reflectors (SLRs) to build cavities. Fabry-P\'erot cavities constructed from these fabrication-tolerant SLRs achieve an intrinsic quality factor of $2 \times 10^6$. By implementing thermal isolation trenches, we also demonstrate a $>10\times$ reduction in power consumption for thermo-optic phase shifters, achieving a $\pi$-phase shift ($P_\pi$) with just 2.5 mW. These tunable and efficient components are key for scaling up to complex photonic integrated circuits.

\end{abstract}

\maketitle

\section{Introduction}
\begin{figure*}[ht!]
\centering
\includegraphics[width=\linewidth]{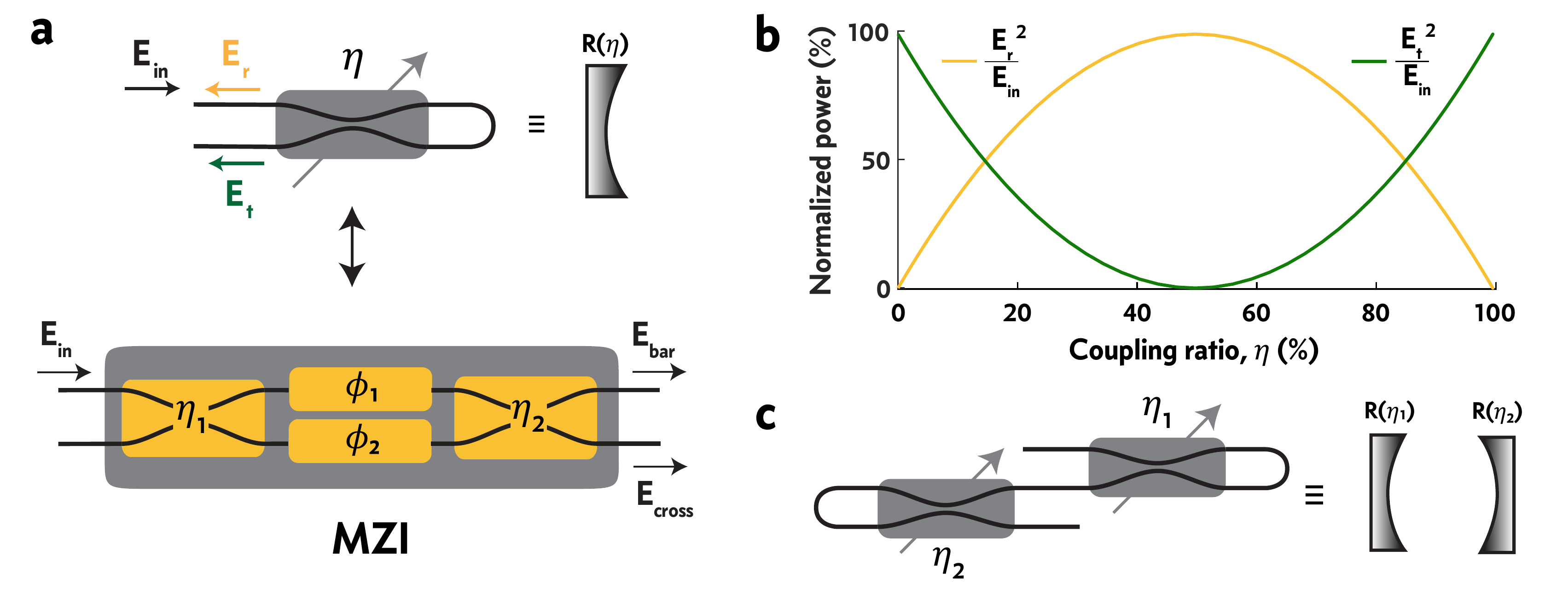}
\caption{Concept of tunable Sagnac loop reflectors with MZIs. a) Schematic of an SLR with a variable beamsplitter. Equivalent to a mirror with a reflectivity parameterized by $\eta$. Variable beamsplitter is achieved with an MZI, illustrated below. b) Reflected and transmitted power of an SLR as a function of beamsplitter ratio $\eta$. c) Fabry–Pérot cavity formed by SLRs with tunable beamsplitters.}
\label{fig:fig1}
\end{figure*}

As photonic integrated circuits (PICs) evolve toward greater complexity and functionality, there is an increasing demand for components that combine ultra-low loss, high tunability, and robustness to fabrication imperfections. Such components must be easily cascaded, allowing scalable construction of programmable photonic networks, including optical neural networks, quantum information processors, and neuromorphic computing architectures~\cite{bogaerts2020programmable, harris2018linear, shen2017deep, wright2022deep, wanjura2024fully, rudolph2017optimistic, valdez2025highcontrastnullingphotonic, miller2025universalprogrammableselfconfiguringoptical}. In a similar vein to classical error-correction where redundancy is used to improve performance, increasing the number of tunable components in a PIC has been shown to extract nearly optimal functionality from suboptimal elements~\cite{Miller:15, Wilkes:16, miller2013self, Miller:13, hamerly2022asymptotically, hamerly2024towards, Bandyopadhyay:21}. 

Mach-Zehnder interferometers (MZIs) are commonly employed as essential units within these systems due to their ability to dynamically tune both the amplitude and phase of propagating light. MZIs however add losses primarily from beamsplitter inefficiencies, metal-induced absorption in active tuning elements, and waveguide scattering losses. Minimizing and accurately characterizing these losses is particularly critical in quantum nonlinear photonics, where the device quality directly limits system performance~\cite{park2024single, stokowski2023integrated}. Single-pass loss measurements of MZIs may not be precise in the low-loss limit because calibration uncertainties can dominate. Cascading multiple devices can improve the sensitivity to loss, however they are experimentally complicated and suffer from compounded fabrication uncertainties~\cite{Desiatov:19}. In contrast, resonant measurements provide intrinsic calibration by isolating the component within a resonator, making the measurements self-consistent and significantly more reliable and sensitive in the low-loss limit.

In this work, we introduce a resonant method to characterize the intrinsic loss of MZIs with minimal experimental complexity. We build MZIs into a Sagnac loop reflector (SLR) configuration and form Fabry–Pérot cavities to perform precise measurements of the MZI-induced losses~\cite{Qiao:25}. Whereas previous approaches in which MZIs are used to tune side-coupling into optical resonators measure extinction ratio, our approach uses an operating point that is inherently robust against beamsplitter imperfections~\cite{Herrmann:24, hou2023programmable} and therefore directly measures loss. This leads to a significantly higher tolerance to fabrication imperfections.

In addition, we show substantially improved thermo-optic phase shifter (TOPS), which are widely used due to their material compatibility and DC-stability~\cite{Harris:14}. Traditional TOPS are difficult to scale up due to their large static power dissipation and crosstalk to other areas of the chip. By incorporating substrate-underetched air trenches for thermal isolation, we achieve over an order of magnitude reduction in the power required for a $\pi$-phase shift ($P_\pi$), from 80 mW down to 2.5 mW~\cite{vanNiekerk:22, Dong:10, ying2019thermally, doi:10.1126/science.ade8450, Liu:20}.  This paves the way for integrating a significantly larger number of tuned components on a single photonic chip.

Although large-scale photonic circuit demonstrations have been dominated by silicon photonics, the advantages of thin-film lithium niobate (TFLN) for nonlinear optics are starting to be implemented in experiment~\cite{zhu2021integrated, hu2025integrated}.  Here, by showing highly tunable, robust, and ultra-low-loss MZI-based reflectors and cavities on TFLN, we demonstrate the feasibility of constructing sophisticated resonant and nonlinear photonic circuits.

\section{Theory and Design}

\subsection{Sagnac loop reflector}
We model an SLR by transferring the optical fields through each component illustrated in Figure~\ref{fig:fig1}(a). There is a beamsplitter with an optical power coupling \(\eta\),  a loop region with a phase shift of $e^{i \phi_\text{SLR}}$ and a power loss of $\gamma_\text{SLR}$. The reflected optical field, derived in Supplemental Document, can be written as
\begin{equation}
\frac{E_\text{r}}{E_\text{in}} = 2i\sqrt{\eta}\sqrt{1-\eta}\sqrt{1-\gamma_\text{SLR}}e^{i\phi_\text{SLR}} \equiv R(\eta). \label{eq:ref}
\end{equation}

From~\eqref{eq:ref}, we see that the reflected field from this loop reflector is a function of the beamsplitter ratio, $\eta$. Being able to tune the coupling ratio $\eta$ effectively leads to a reflector with a tunable reflectivity as illustrated in Figure~\ref{fig:fig1}a. The tuning curve of this SLR is shown in Figure~\ref{fig:fig1}b. Note that at 50\% beamsplitter coupling ratio, there is perfect reflection for a lossless SLR and is partially reflective for coupling ratios close to 50\% (falls off with second order in $\eta$).

Placing two of these components back to back forms a Fabry–Pérot resonator with independently tunable out-couplings through each reflector (Fig~\ref{fig:fig1}c). If the waveguide connecting the two components is short, the losses of this cavity will be dominated by the components in the tunable reflector. 

\begin{figure*}[ht!]
\centering\includegraphics[width=\linewidth]{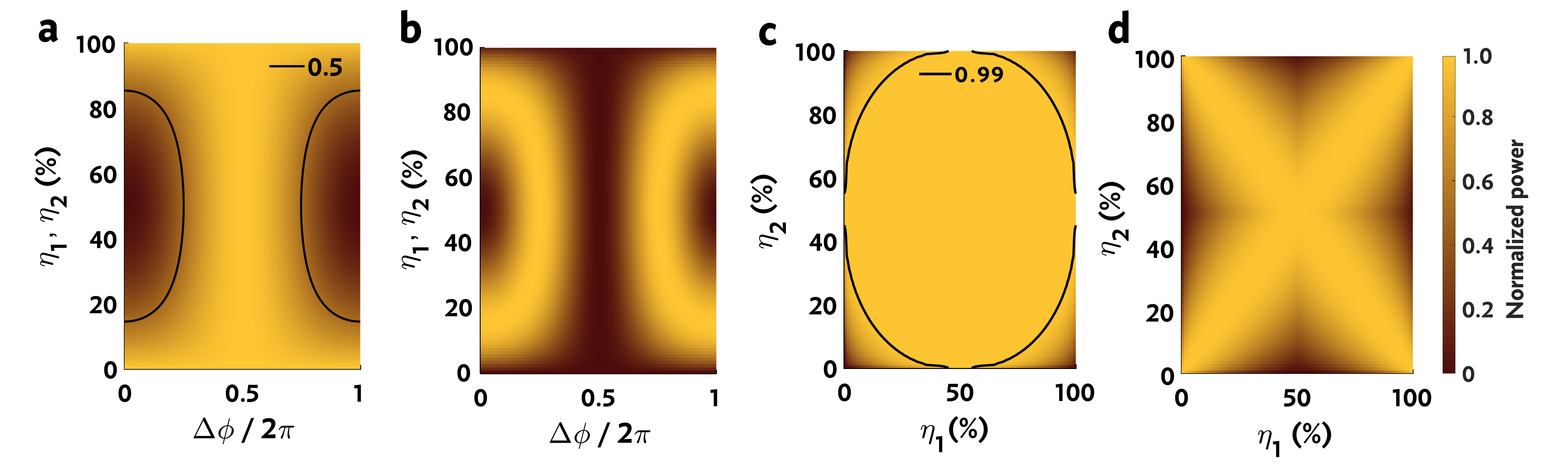}
\caption{Simulated MZI outputs and SLR reflectivities. a) Bar output of a balanced MZI at various beamsplitter couplings and phase shifts. b) Reflection of an SLR controlled by a balanced MZI. c) Maximum achievable reflection of an SLR with two different couplers in the MZI. d) Maximum achievable transmission through the SLR.}

\label{fig:fig2}
\end{figure*}

\subsection{Mach-Zehnder interferometer}
To build the tunable reflector, we require a variable beamsplitter, for which we use a tunable MZI, as illustrated in the lower box in Figure~\ref{fig:fig1}a. We model an MZI as two beamsplitters with power splitting ratios $\eta_1, \eta_2$, each with losses $\gamma_1, \gamma_2$, and a phase shift in each arm of $\phi_1, \phi_2$. We only consider symmetric MZIs, where the optical path length in each arm is the same. Thus the relative phase shift between the two arms is caused solely by the active component, \textit{i.e.}, the phase shifter. The field transmission at the ``through'' (or ``bar'') output port of the MZI can be written as
\begin{align*}
\frac{E_\text{bar}}{E_\text{in}} = 
\left[ 
\sqrt{1 - \eta_1} \sqrt{1 - \eta_2} e^{i\phi_1} 
- 
\sqrt{\eta_1} \sqrt{\eta_2} e^{i\phi_2} 
\right] \\\times
\sqrt{1 - \gamma_1} \sqrt{1 - \gamma_2}.
\end{align*}
We plot $|{E_\text{bar}}/{E_\text{in}}|^2$ for an MZI with identical beamsplitters ($\eta_1 = \eta_2$) at various coupling ratios and phase differences between each arm ($\Delta \phi = \phi_1-\phi_2$) in Figure \ref{fig:fig2}a. The dark regions indicate that \textit{high extinction} on the bar port is only achieved when the couplers approach perfect 50:50 splitting so that the fields can be canceled perfectly on one port. In contrast, to construct a highly reflective Sagnac reflector, the important requirement is only to have the MZI outputs split by 50:50, shown as the 0.5 contour line in Fig~\ref{fig:fig2}a . We plot the reflectivity of a Sagnac loop reflector, $|{E_\text{r}}/{E_\text{in}}|^2$, formed by connecting the two outputs of the MZI, shown in Figure~\ref{fig:fig2}b. As expected, we see that beamsplitter coupling ratios within the MZI spanning 15:85 to 85:15 can still result in a highly reflective reflector in a balanced MZI design, evident from the bright regions in Figure~\ref{fig:fig2}b. This relaxes the requirement to have perfect 50:50 beamsplitter coupling ratios in an MZI to achieve high extinction.

\subsection{Unequal beamsplitters}
The 50:50 coupling requirement for a highly reflective loop reflector also relaxes the requirement to have identical beamsplitters in an MZI. The outputs of an SLR with unbalanced MZI beamsplitters ($\eta_1 \neq \eta_2$) are illustrated in Figures \ref{fig:fig2}c and d. In this analysis, we consider each possible pair of coupling ratios $\eta_1, \eta_2$, and the maximum reflected and transmitted power is calculated by sweeping the MZI phase $\Delta \phi$. We plot the maximum reflected and transmitted power in Figs~\ref{fig:fig2}c and d, respectively. Fig~\ref{fig:fig2}c shows that high reflection from these MZI-controlled SLRs can be achieved for a wide range of beamsplitter ratios (99\% reflection on the black contour line), showing strong robustness to fabrication variations in the MZI beamsplitters. This is highly advantageous for applications that require broadband reflectivity, eg. integrated lasers~\cite{zhang2014sagnac, Yu:23}. 

Conversely, realizing the highly transmissive state (low reflectivity) requires the output ports of the MZI to be 0:100 or 100:0, which is more sensitive to coupler imbalance. High transmission through the reflector is plotted as bright regions in Fig~\ref{fig:fig2}d. To understand this behavior, we analyze the maximum bar and cross ports of a conventional lossless MZI with unbalanced beamsplitters, results shown in Supplemental Document. To get complete power transfer to the bar port of the MZI (100:0), one must have identical beamsplitters ($\eta_1 = \eta_2$). To get all the power into the cross port (0:100), the two beamsplitters must sum to unity ($\eta_1 + \eta_2 = 1$). Thus, a fully transmissive Sagnac loop reflector can be achieved under these two conditions.

For a conventional MZI, achieving maximum power at the two outputs using a single tunable phase shifter requires both beamsplitters to be precisely 50:50. In contrast, the analysis here shows that achieving both total reflection and transmission in a single MZI-controlled SLR is possible with identical beamsplitters with any ratio between 15:85 to 85:15. This increased tolerance to splitting ratio ultimately arises because light travels through the MZI twice due to the loopback, which is analogous to compensating for imperfections by cascading two MZIs in succession~\cite{Wilkes:16}. 

\section{Experimental Results}

\subsection{Fabrication} \label{sec:fab}

\begin{figure}[ht!]
\centering\includegraphics[width=\linewidth]{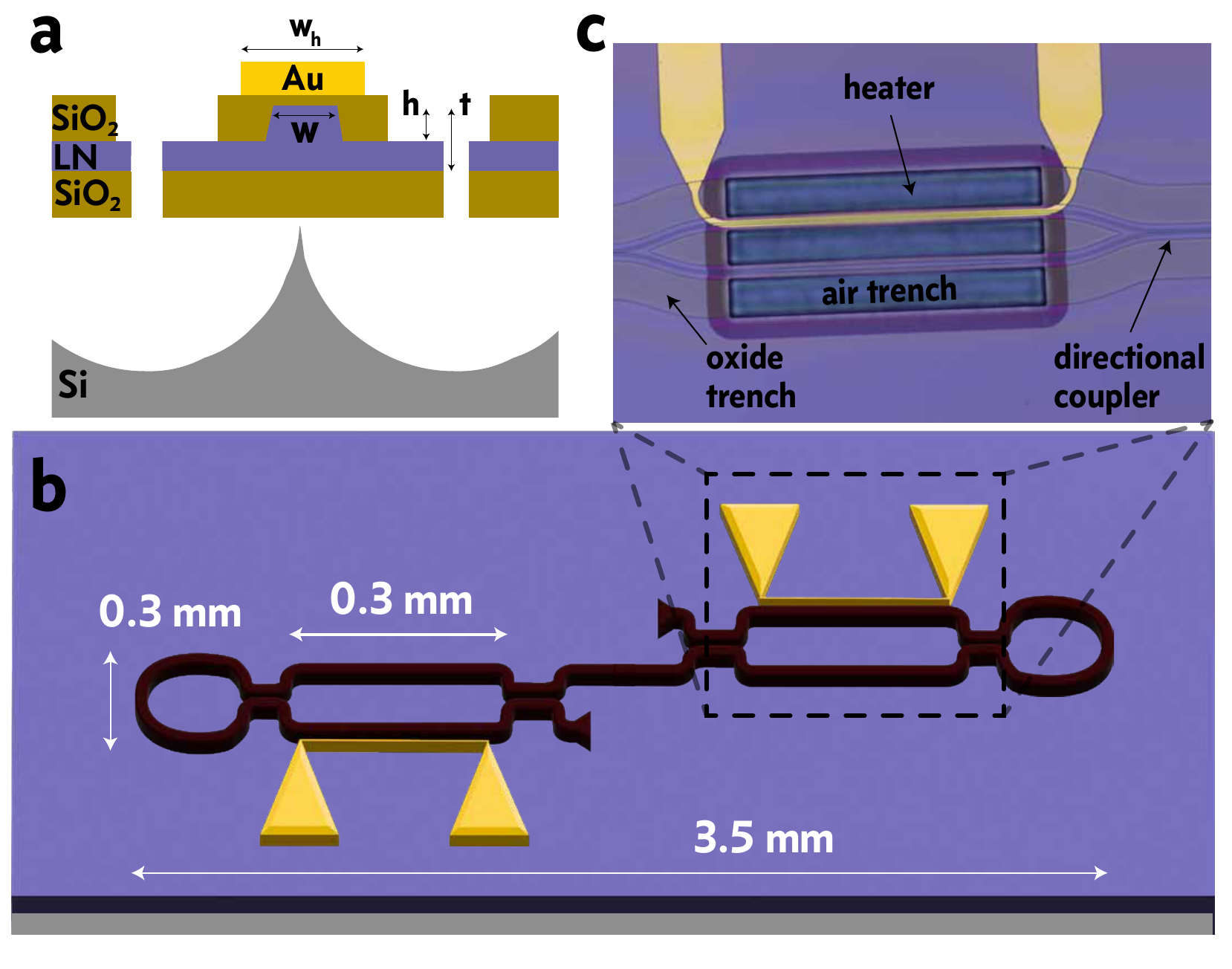}
\caption{Tunable Sagnac Fabry–Pérot cavities on LNOI. (a) Photonic stack, $w_\text{h} = 5~\mu$m, $w = 1.2~\mu$m, $h = 300$~nm, $t = 500$~nm (b) Illustration of device layout (not to scale) (c) Optical microscope image of the thermo-optic phase shifter.}
\label{fig:fig3}
\end{figure}

\begin{figure*}[ht!]
\centering\includegraphics[width=\linewidth]{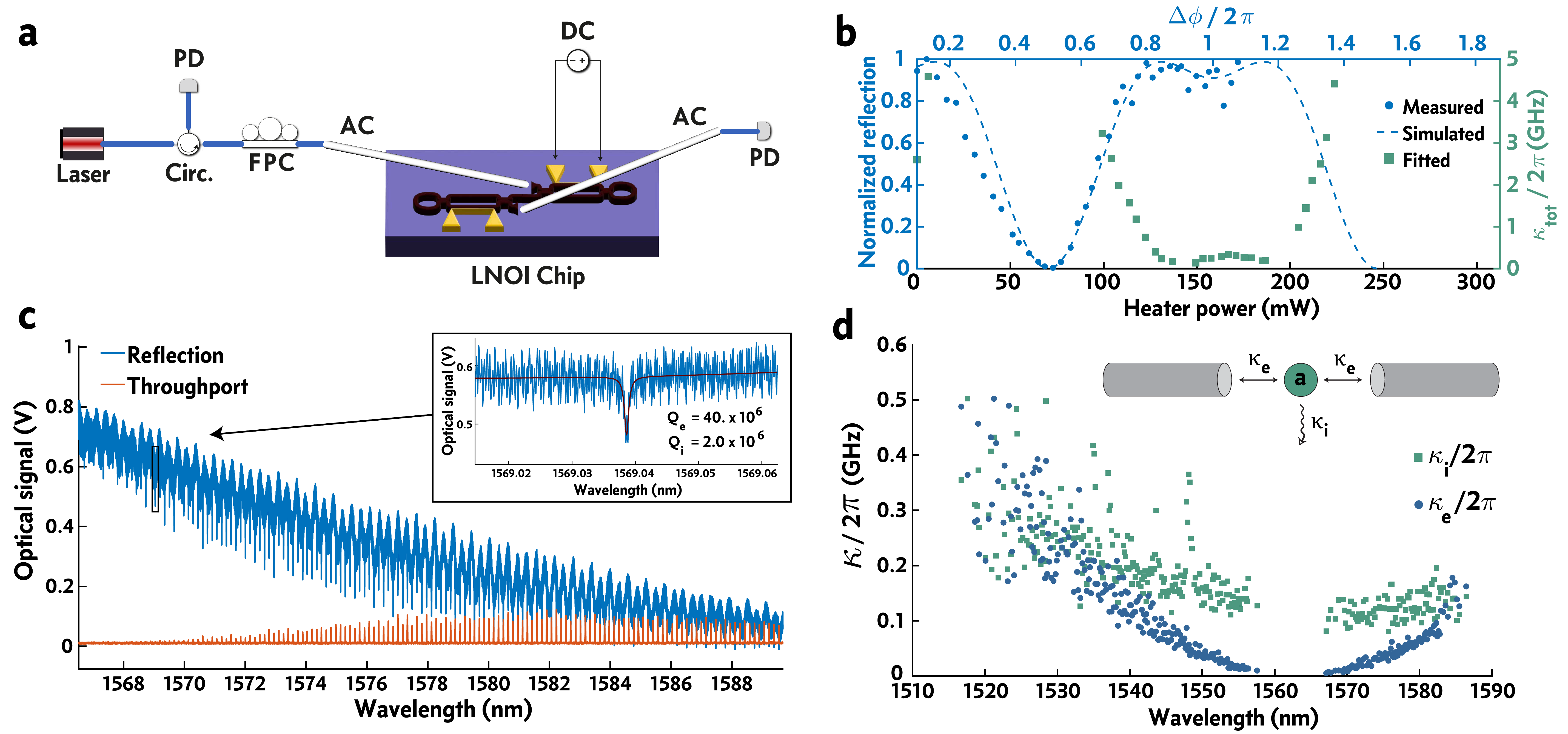}
\caption{Characterization of tunable SLRs and Fabry–Pérot cavities. a) Measurement setup.  b) Tuning curve of SLR and Fabry–Pérot cavity. Measured reflection from a control SLR in blue dots. Simulated reflection of an SLR with 80\% beamsplitters in MZI in dashed blue line. Fitted linewidths of Fabry–Pérot cavity mode in green squares. c) The reflected and transmitted spectrum of the Fabry–Pérot cavity at one highly reflective operating point. d) Fitted modes across the reflected spectrum with double-sided cavity model.
PD: photodetector, FPC: fiber polarization controller, AC: angle-cleaved fibers, DC: direct-current power supply.}
\label{fig:fig4}
\end{figure*}

We fabricate the devices on an X-cut lithium niobate on insulator (LNOI) stack, with a 500 $\mu$m Si handle, 3 $\mu$m buried oxide layer, 500 nm MgO-doped lithium niobate layer, shown in Figure~\ref{fig:fig3}a~\cite{Zhang:17, park2024single}. The MZI beamsplitters are directional couplers with a length of $180\ \mu$m and a gap of $0.9\ \mu$m. We write the waveguides with 100 kV Raith 5200+ electron beam lithography on an HSQ mask. We transfer the patterns to LN with Ar+ ion mill etching, and clean the devices post-etch using an HF chemistry. We clad the waveguides with 1 $\mu$m thick PECVD oxide to avoid metal-induced losses. We add thermo-optic phase shifters to only one arm of the MZI for differential tuning, and these TOPS are fabricated with photolithography on a bilayer LOR5A-SPR3612 resist stack. Cr/Au is chosen as the metal stack due to compatibility with electro-optic devices, but Ti/Pt can also be used due to its resistance to oxidation at high temperatures~\cite{hu2022high}. The resistance of the heater is estimated to be 25 $\Omega$ by using a PLH250 high-voltage DC power supply. 

The device is illustrated in Figure \ref{fig:fig3}b. The total optical path length is about 7 mm while the phase shifter in each MZI is about 0.3 mm. The loop reflectors use Euler bends to minimize the bending loss.

Subsequent substrate undercutting for thermal isolation is done with a $4~\mu\text{m}$ thick SPR220-3 photoresist mask. An Ar+ ion mill is used to etch the lithium niobate slab, CF$_4$/CHF$_3$ is used to etch the oxide layers, and XeF$_2$ is used to isotropically etch the silicon substrate. An optical microscope image of the thermal phase shifter is shown in Figure \ref{fig:fig3}c.

\subsection{SLR reflectivity curve and Fabry–Pérot mode spectrum}

We characterize the mode spectrum of these Fabry–Pérot resonators by measuring reflection and transmission with a setup shown in Figure~\ref{fig:fig4}a. Light is coupled on chip via angle-polished SMF-28 fibers into grating couplers. A telecom fiber circulator is placed before the input fiber to route the reflected light from the first SLR into a Newport Nanosecond Photodetector.

We measure the response from a single SLR and a double SLR Fabry–Pérot cavity as we tune thermal power applied to the MZIs. The reflection of the control SLR, shown as blue dots in Figure~\ref{fig:fig4}b, can be described with our scattering matrix model when the beamsplitters in the MZI are at 80\% power coupling, shown as blue dashed line in Figure~\ref{fig:fig4}b. This is within the 15:85 to 85:15 beamsplitter coupling range for full reflection. From the SLR tuning curve, we see that the amount of heater power to induce a $\pi$ phase-shift in the MZI is around 80 mW.

To characterize the Fabry–Pérot cavity, we tune both SLRs simultaneously and symmetrically by applying equal heater powers and fit the total linewidth of a single mode with a Lorentzian resonance, plotted as green squares in Figure~\ref{fig:fig4}b. We see that the linewidths follow the transmission through the reflector, which is the opposite of the reflector reflection tuning curve. This is expected as a larger transmission through the reflectors means a larger coupling of the cavity mode to the environment. Between 120 and 180 mW of applied power, the cavity linewidths are low due to the high SLR reflectivity, and this region allows us to extract the internal loss of the modes.

We plot the Fabry–Pérot mode spectrum at a high reflectivity point (120 mW heater power) in Figure~\ref{fig:fig4}c, reflection and transmission in blue and orange, respectively. The entire mode spectrum is measured by sweeping the wavelength of a Santec TSL diode laser, and the background response in the reflected signal is due to the wavelength dependence of our grating couplers. The free spectral range, $f_\text{FSR}$ for this mode family is 19 GHz and the internal quality factor $Q_i$ of one of the modes is $2.0 \times 10^{6}$, fitted with a symmetric double-sided bus-coupled cavity Lorentzian model and shown in the inset of Figure~\ref{fig:fig4}c. The external quality factor $Q_e$ is found to be $40 \times 10^{6}$, indicating the cavity is very weakly coupled to the environment due to the high reflectivity of the SLRs.

The modes across the reflected spectrum in Fig~\ref{fig:fig4}c are fit with the same double-sided Lorentzian model and plotted in Figure~\ref{fig:fig4}d. We see that the internal loss rate stays pretty constant around 150 MHz between the region of 1540 and 1580 nm, while the external coupling rate dips way down and becomes almost zero from 1560 to 1570 nm. This wavelength dependence of the cavity SLRs can be attributed to the wavelength dependence of the two directional couplers in the MZIs, giving us a bandwidth of around 10 nm (around 1 THz).

\subsection{Extracting loss from quality factor} \label{sec:ds}

We use a symmetric double-sided bus-coupled cavity model to fit our Fabry–Pérot modes because we tune both SLRs equally in order to keep the external coupling in each mirror the same. This model is written as,
$$
\left|\frac{\alpha_\text{out}}{ \alpha_\text{in}}\right|^2 = \frac{4\Delta^2 + \kappa_i^2}{4\Delta^2 + (\kappa_i+2\kappa_e)^2}
$$
where $|\alpha|^2$ is photon number flux, $\Delta$ is the detuning from resonance, $\kappa_i$ is the internal linewidth, and $\kappa_e$ is the external coupling rate. The Fabry–Pérot mode we found had an internal quality factor of $Q_{i_\text{FP}} = \omega_0/\kappa_i = 2 \times 10^6$, from which we can extract the round trip loss using the following relation,
$$
l = \frac{2\pi f_0}{f_{\text{FSR}} \cdot Q_i}.
$$
We find $l = 3\%$. If we attribute all this loss due to the two MZIs within the Fabry–Pérot cavity, this gives us a loss of 1.5\% per MZI (-0.07 dB).

In actuality this is a conservative estimate of MZI loss because this value includes all the potential loss mechanisms within the Fabry–Pérot cavity, which we can sort into to waveguide propagation loss and MZI losses. Waveguide propagation loss and quality factor are related by,
 \[
\alpha = 4.34 \cdot \frac{2\pi n_g}{Q \lambda_0}\quad [\text{dB m}^{-1}].
\]

State-of-the-art LNOI fabrication has pushed waveguide propagation losses as low as 3 dB/m~\cite{Desiatov:19, Zhu:24}. Using the low-loss value of 3 dB/m and a group index of $n_g = 2.26$, we find that propagation losses would result in a quality factor of $Q_{i_\text{propagation}} = 13 \times 10^6$.

We may subtract out the contribution from propagation loss from our measured quality factor as follows
$$
\frac{1}{Q_{i_\text{FP}}} = \frac{1}{Q_{i_\text{propagation}}} + \frac{2}{Q_{i_\text{MZI}}}
$$
note that the factor of 2 comes from the fact that we have two MZIs in the cavity. 

We find that $Q_{i_\text{MZI}} = 5 \times 10^6$ corresponding to a round trip loss of 1.3\% (-0.06 dB per MZI). The MZIs only had a path length of 1 mm, so the extra sources of loss comes from the two directional couplers and metal-induced loss from the thermo-optic phase shifter. However, we do not measure the actual propagation loss of our device so the best we can claim is the conservative loss estimate of -0.07 dB per MZI. Future experiments should include test resonators for extracting propagation loss.

As seen from the fitted modes plotted in Figure~\ref{fig:fig4}d, the largest external quality factor that was able to be measured was $40 \times 10^{6}$, leading to a power coupling of 0.2\% through each SLR. Some modes were too undercoupled to be measured, seen as the gap from 1560 to 1565 nm. This measurement shows that the transmission through the reflector can be tuned below -27 dB, indicating it is possible to obtain high extinction with a Sagnac loop reflector even with suboptimal beamsplitter design. To be sensitive beyond -27 dB of coupling would require a lower internal loss in order to observe resonances.

\section{Power-efficient thermo-optic tuning}
\begin{figure}[ht!]
\centering\includegraphics[width=\linewidth]{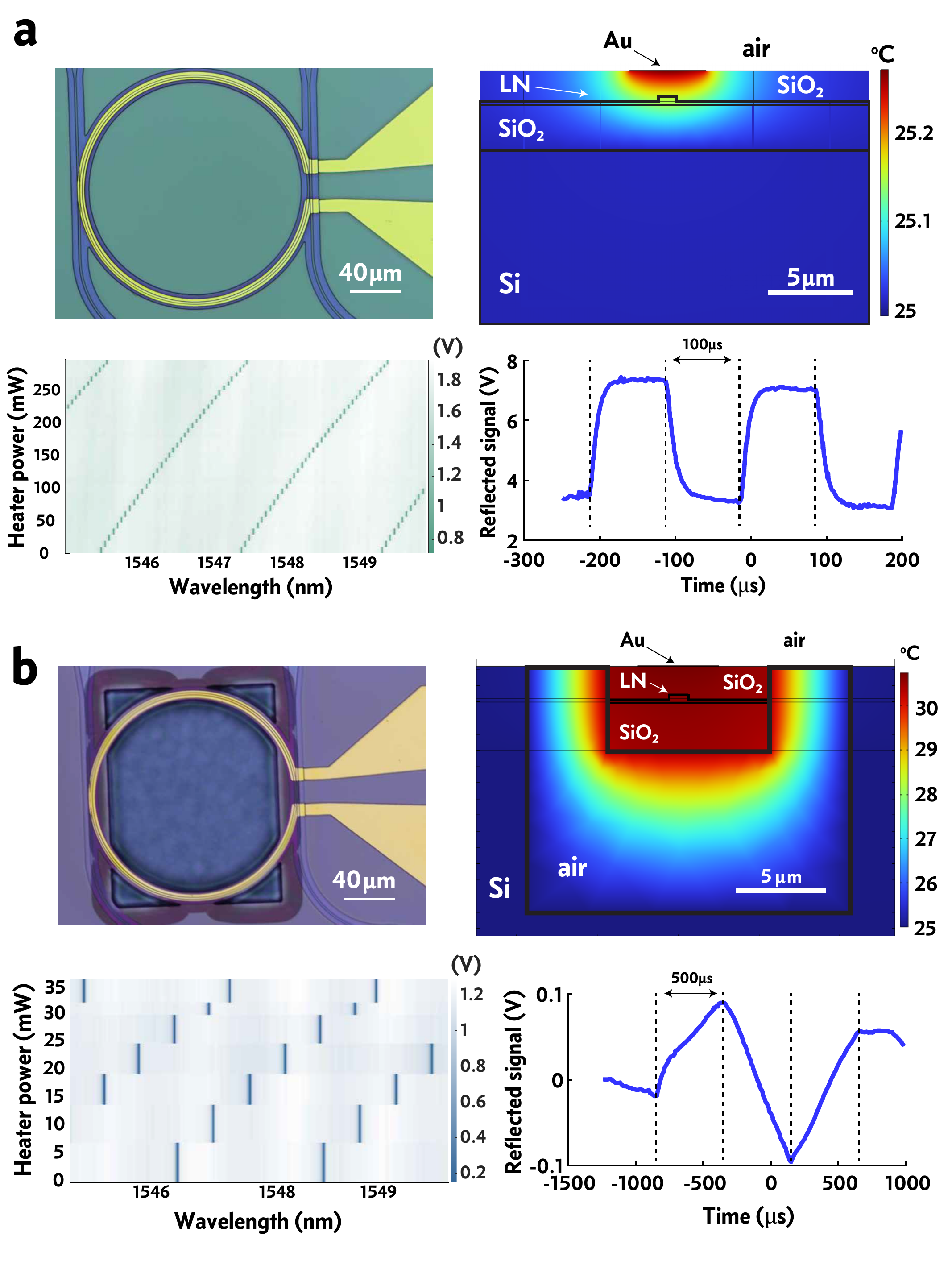}
\caption{TOPS characterization with (a) normal heater (b) undercut heater. Top left plot: microscope image of ring resonator. Top right plot: COMSOL simulation of heat flow. Bottom left plot: transmission of ring resonator as electrical power is applied to the heaters. Bottom right plot: reflected signal of an SLR with a square wave applied to the MZI phase shifters.}
\label{fig:fig5}
\end{figure}

\subsection{Ring resonator}

\begin{figure*}[ht!]
\centering\includegraphics[width=\linewidth]{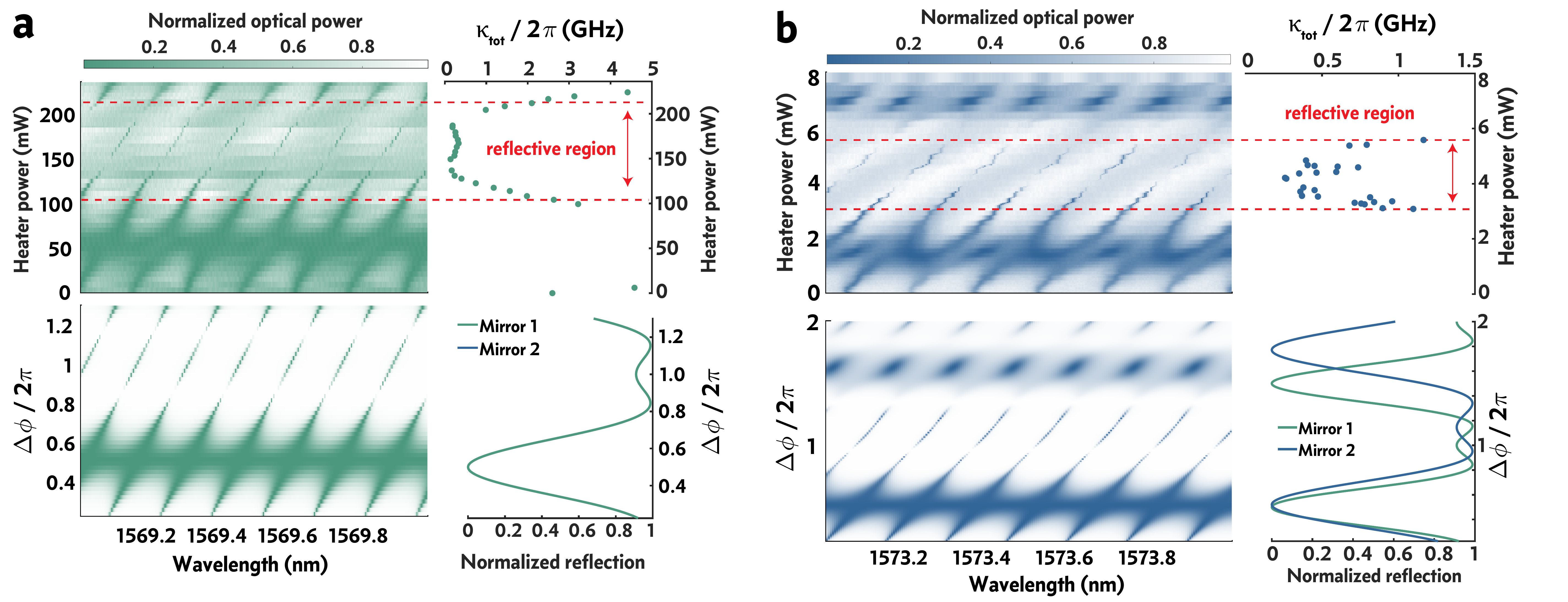} 
\caption{Full mode tuning of the Fabry–Pérot resonators (a) before undercut (b) after undercut. Top left: measured experimental data. Bottom left: simulated scattering matrix model. Top right: fitted linewidth of a single mode. Bottom right: reflector reflectivity curve used in scattering matrix model. }
\label{fig:fig6}
\end{figure*}

\begin{table}[htbp]
\caption{Thermal Conductivity of Various Materials~\cite{multiphysics1998introduction}}
\label{tab:thermal-conductivity}
\centering
\begin{tabular}{lc}
\hline
Material & Thermal Conductivity (W/m·K) \\
\hline
Air              & 0.02  \\
Silicon          & 130   \\
Silica           & 1.38  \\
Lithium Niobate  & 4.6   \\
Gold             & 317   \\
Platinum         & 71.6  \\
\hline
\label{tab:therm}
\end{tabular}
\end{table}

We use ring resonators to test the thermo-optic tuning strength of our metal heaters, top left images in Figure~\ref{fig:fig5}. We perform mode spectroscopy of the ring resonator as we apply power through the heaters, plotted in bottom left in Figure~\ref{fig:fig5}. For a normal ring with a heater placed on top of the cladded waveguide, the mode redshifts with a rate of 6 pm/mW, which agrees well with literature~\cite{Liu:20}. To increase the efficiency of thermo-optic tuning, we undercut the waveguides by substrate etching, detailed in section~\ref{sec:fab}. We see that the tuning rate increases to around 70 pm/mW, over an order of magnitude increase in efficiency. Since the 3\ $\mu$m buried oxide layer is not affected by the silicon etch, we expect the waveguide mode index to remain the same. The quality factor of the rings were not sensitive to the substrate undercut process and the devices were mechanically stable due to tethering at four points. 

To understand thermal isolation in our device, we simulate a cross-section of our LNOI stack using Heat Transfer in COMSOL Multiphysics, shown in the top right plots in Figure~\ref{fig:fig5}. The thermal conductivities of each material are listed in table~\ref{tab:therm}. We see that applying 1 mW of power to the heater locally increases the waveguide temperature by 0.2\ C. Due to the high thermal conductivity of silicon, the top of the silicon is in equilibrium with the bottom platform, which is set to 25\ C with a copper stage and a thermo-electric cooler. Without any isolation, the applied heater power quickly thermalizes down to the silicon temperature. 

Using air isolation improves the localization of heat. The high thermal resistivity of an air trench creates a large temperature gradient allowing for the temperature near the waveguide to be remain very high. Now 1 mW of heater power increases the waveguide temperature by 5\ C, about a 25 times stronger tuning-rate which also agrees with previous literature~\cite{Liu:20}. 

We also measure the bandwidth of our TOPS by applying a square wave to the MZI with a RIGOL function generator and monitoring the reflection of a SLR device with a Tektronix oscilloscope, plotted in the bottom of Figure~\ref{fig:fig5}. We modulated the SLR before air isolation with a 10 kHz square wave and after air isolation with a 2 kHz wave. Our results show that the radio-frequency bandwidth of our TOPS decreased by over an order of magnitude with air isolation, in agreement with previous literature. Part of decreased bandwidth is attributed to the increased thermal load of our device, and part of it is also attributed to the long length of the trench and the lack of tethers, causing mechanical instability. This can be avoided in the future by substrate etching through windows that are staggered on either side of the waveguide.

\subsection{Fabry–Pérot resonator}

We characterize the full parameter space of the tunable Sagnac Fabry–Pérot resonators by performing mode spectroscopy while applying equal heater power to both of the reflectors. The results are plotted on the top left in Figure~\ref{fig:fig6}, normal in a, and substrate undercut in b. When the SLR reflectivities are low, the reflected signal from the Fabry–Pérot cavities is also low. Narrow linewidths form when the SLR reflectivies are high, and we get sharp dips in reflection which are the cavity resonances. 

We track a single mode with various applied phase shifts and plot the linewidths in the top right plots in Figure~\ref{fig:fig6}. The labeled ``highly reflective'' region leads to sharp cavity modes, and this region decreases from 100-200 mW down to 3-6 mW with thermal isolation. We also simulate this resonator tuning with our numerical scattering matrix model, plotted on the bottom left in Figure~\ref{fig:fig6}. Our  model can explain the behavior of the Fabry–Pérot cavity response as the linewidths are tuned, meaning that we have captured most of the physics of the system. The reflectivity of each SLR is a function of the phase shift in the MZI, which is a tunable parameter in our model. To get qualitative agreement with our measurements, the two reflectors must be slightly offset after the undercut, as seen in the bottom right plot in Figure~\ref{fig:fig6}b. The phase of one loop reflector $\phi_1$ tunes slightly differently with the same applied voltage than the phase of of the second loop reflector $\phi_2$, related by $\phi_2 = 0.8\cdot\phi_1 + \pi/6$. We attribute this asymmetry to the mechanical instability of the undercut process, as the local heating may cause the suspended waveguides to bend. This may be improved in the future with tethers to stabilize the waveguides. Our results show that we can fully control the cavity linewidths with under 10 mW of heater power using air-isolated TOPS.

\section{Discussion}
These MZI-controlled Sagnac loop reflector Fabry–Pérot cavities allows one to measure both the loss and the bandwidth of an MZI with minimal added complexity. We realize that the Sagnac loop reflector parameter space is robust to beamsplitter design, and therefore this method would work for most MZI devices. Furthermore, these SLR devices could also work for most integrated waveguide-based photonic platforms, as the tuning relies soley on the thermo-optic effect. The low-loss, large tunability, and robustness to fabrication errors make MZI-controlled SLRs a useful component for photonic integrated circuits~\cite{ahmed2025universal, aghaee2025scaling}.

On TFLN, we show here the loss is quite low (-0.07 dB per MZI), meaning that MZIs can be cascaded without significant loss. Further reductions in loss could be attained with an optimized cladding recipe, chemical clean processes, and metal layout. Our measurements also show that tunable Sagnac loop reflectors with MZIs can be almost a perfect reflector with a leakage below -27 dB and they can be used to form high quality factor Fabry–Pérot cavities. The bandwidth of these reflectors are quite high (THz) and limited by the dispersion of the directional couplers within the MZI. Broadband directional couplers may further increase this bandwidth~\cite{Lu:15}

Finally we have shown that TOPS can be operated at milliwatt of electrical power with proper silicon substrate etching. The optical loss increased due to improper mechanical stability and future work should focus on tethering waveguides to improve stability and minimize loss. Air trench isolated TOPS have a low temporal bandwidth, but for slow DC tuning this is an optimal approach compared to electro-optic phase shifters (EOPS), which can suffer from bias drifts~\cite{celik2024roles}. 

Low-loss and highly-tunable MZIs and Sagnac loop reflectors are fundamental building blocks in large-scale programmable photonic circuits~\cite{Jiang:16, park2024fully, Yu:23, Lepert_2013,el2022modelling, zhang2014sagnac, sun2013tunable, arianfard2023sagnac}. These components may become crucial in applications requiring many coupled resonant modes, such as programmable neuromorphic computing networks~\cite{wanjura2024fully}.

\begin{acknowledgments}
The authors would like to acknowledge Felix Mayor, Rachel Gruenke-Freudenstein, Gitanjali Multani, Wentao Jiang, Oğuz Tolga Çelik, and Nancy Yousry Ammar for helpful discussions and advice. The authors wish to thank NTT Research for their financial and technical support. This work was supported by the U.S. government through the Defense Advanced Research Projects Agency (DARPA) INSPIRED program, the National Science Foundation NSF-SNSF MOLINO project No. ECCS-2402483, and the US Department of Energy through grant no. DE-AC02-76SF00515 and via the Q-NEXT Center.
Device fabrication was performed at the Stanford Nano Shared Facilities (SNSF) and the Stanford Nanofabrication Facility (SNF), supported by NSF award ECCS-2026822.
L.Q. gratefully acknowledges support from the Shoucheng Zhang Graduate Fellowship Program. This material is based upon work supported by the National Science Foundation Graduate Research Fellowship Program under Grant No. DGE-1656518.
\end{acknowledgments}

\nocite{*}

\bibliography{main}

\clearpage
\onecolumngrid
\section*{Supplemental Information}

\section{Optical field matrix model}
We model light through our components classically with the complex amplitudes of the optical field, denoted by $E$~\cite{buscaino2020design}. We track both the phase and magnitude of the optical field as they go through each component in our MZI-controlled SLR Fabry–Pérot cavities. 

Our device is static in time so the implicit $e^{i\omega t}$ dependence for a monochromatic beam of light can be dropped. Propagation can be seen as a phase shift in the field given by $e^{i\beta(\omega)L}$, where $\beta(\omega)$ is given by the dispersion of our TFLN waveguide.

\subsection{Beamsplitter}
A beamsplitter is a two port linear element, illustrated in Figure~\ref{fig:SI1}(a). We use a symmetric form of the beamsplitter matrix that reads as follows~\cite{loudon2000quantum},

\[
\begin{pmatrix}
E_3 \\
E_4
\end{pmatrix}
=
\begin{pmatrix}
\sqrt{1 - \eta} & i\sqrt{\eta} \\
 i\sqrt{\eta} & \sqrt{1 - \eta}
\end{pmatrix}
\begin{pmatrix}
E_1 \\
E_2
\end{pmatrix}
=
\begin{pmatrix}
S_{31} & S_{32} \\
S_{41} & S_{42}
\end{pmatrix}
\begin{pmatrix}
E_1 \\
E_2
\end{pmatrix}
\equiv
S_{\text{BS}}
\begin{pmatrix}
E_1 \\
E_2
\end{pmatrix}
\]
where $\eta$ is the optical power coupling ratio across the beamsplitter, $E_1$ and $E_2$ are the fields in the upper and lower waveguides respectively.

We model the loss of the beamsplitter classically by including an additional term $\sqrt{1-\gamma}$, where $\gamma$ is the fraction of the optical power lost. The final beamsplitter matrix reads as, 

\begin{equation}
\boxed{
S_{\text{BS}} =
\begin{pmatrix}
\sqrt{1 - \eta} & i\sqrt{\eta} \\
 i\sqrt{\eta} & \sqrt{1 - \eta}
\end{pmatrix}
\sqrt{1 - \gamma} 
}
\label{eq:bs}
\end{equation}

Note that for a lossless beamsplitter ($\gamma = 0$), the above matrix is unitary $S^\dagger S = SS^\dagger = I$ and also the system is reciprocal as expected, $S = S^T$.

\subsection{Mach-Zehnder interferometer}
\begin{figure}[ht!]
\centering\includegraphics[width=\linewidth]{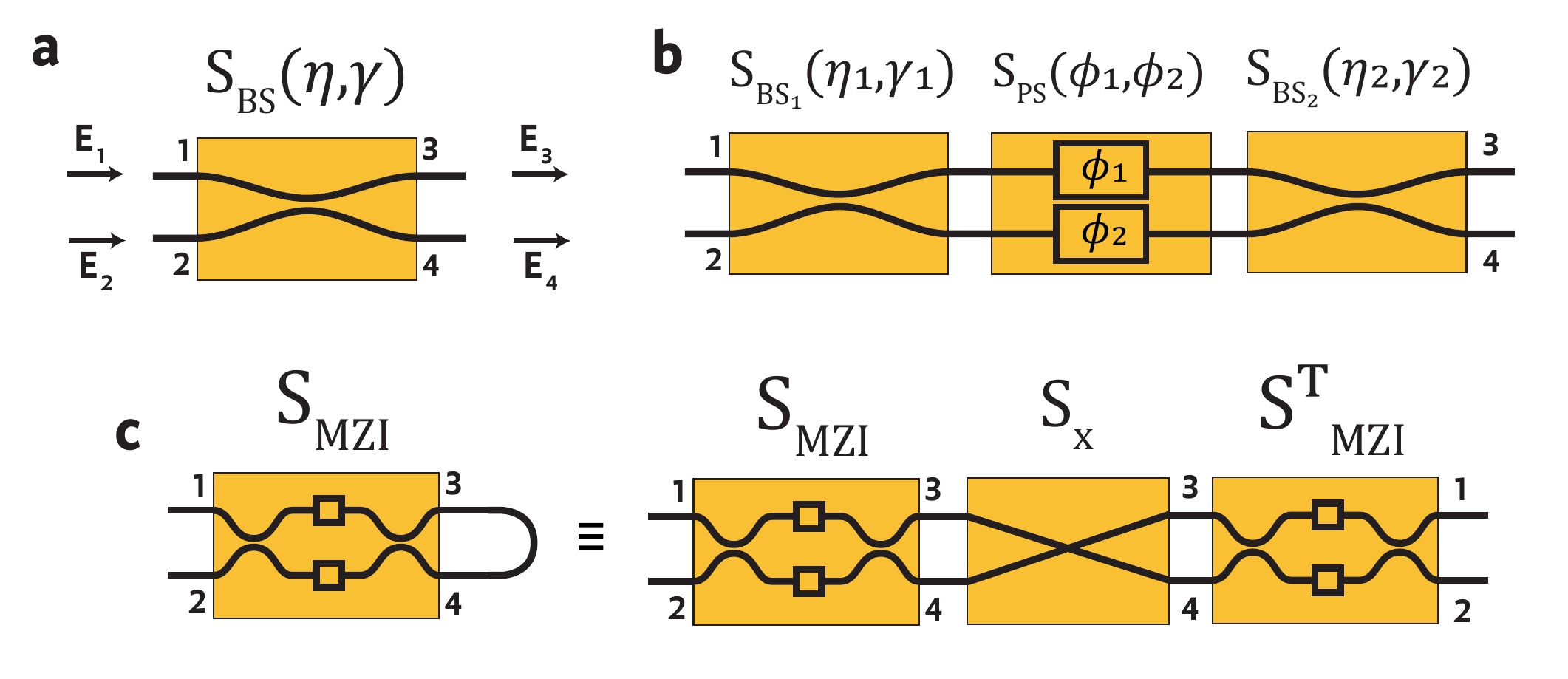} 
\caption{Model methodology. (a) Beamsplitter 2x2 element. (b) MZI as a cascade of beamsplitters and phase shifters. (c) MZI controlled Sagnac loop reflector. The loop can be unraveled to form an equivalent forward propagating picture (on the right).}
\label{fig:SI1}
\end{figure}

Now the MZI in Figure~\ref{fig:SI1}(b) consists of two beamsplitters and a phase shifter in each arm. The phase shift in each arm is generically given by, $\phi = \beta(\omega) L + \phi_{ps}$, where $\beta(\omega)L$ is due to propagation and $\phi_{ps}$ is the additional phase of a controllable phase shifter. Since all our MZIs are path length matched, we can drop the first term on the right hand side and only keep the phase shift due to an active phase shifter, $\phi_{ps}$.

Each arm can have different phase shifts, denoted by $\phi_1$ and $\phi_2$ in Figure~\ref{fig:SI1}(b). The propagation matrix for the upper and lower arms are described by the following matrix,

\[
S_{\text{PS}} =
\begin{pmatrix}
 e^{i\phi_1} & 0 \\
0 & e^{i\phi_2}
\end{pmatrix}
\]

The MZI scattering matrix is just the product of each component,
$
\boxed{S_{\text{MZI}} = S_{BS_2} S_{PS} S_{BS_1}}
$, and thus the full MZI matrix is

\[
S_{\text{MZI}} =
\begin{pmatrix}
\sqrt{1 - \eta_1} \sqrt{1 - \eta_2} e^{i\phi_1}
- \sqrt{\eta_1} \sqrt{\eta_2} e^{i\phi_2}
&
i \sqrt{\eta_1} \sqrt{1 - \eta_2} e^{i\phi_1}
+ i \sqrt{1 - \eta_1} \sqrt{\eta_2} e^{i\phi_2}

 \\[10pt]

i \sqrt{1 - \eta_1} \sqrt{\eta_2} e^{i\phi_1}
+ i \sqrt{\eta_1} \sqrt{1 - \eta_2} e^{i\phi_2}
&
- \sqrt{\eta_1} \sqrt{\eta_2} e^{i\phi_1}
+ \sqrt{1 - \eta_1} \sqrt{1 - \eta_2} e^{i\phi_2}
\end{pmatrix}
\sqrt{1 - \gamma_1} \sqrt{1 - \gamma_2}
\]
where $\gamma_1, \gamma_2$ and the losses in the first and second beamsplitters respectively. The ``bar'' and ``cross'' ports described in the main text are the $S_{31}$ and $S_{41}$ components of the scattering matrix above. 

Two remarks. First, for a lossless MZI ($\gamma_1 = \gamma_2 = 0$), the two output ports should add to 1 (ie. $ |S_{\text{MZI}_{31}}|^2 + |S_{\text{MZI}_{41}}|^2 = |S_{\text{MZI}_{32}}|^2 + |S_{\text{MZI}_{42}}|^2 = 1$). Second, sending light backwards through the MZI (input to ports 3 and 4) is the same as taking the transpose of the MZI matrix, $S_\text{MZI}^T$.

\subsection{Sagnac loop reflector}
We continue. A Sagnac reflector is essentially a beamsplitter where the two output ports are tied together in a loop, Fig~\ref{fig:SI1}c. We can model a loop with a power loss, $\gamma_\text{SLR}$, phase shift $\phi_\text{SLR} = \beta(\omega)L$, and the bit-flip matrix $\sigma_x$ where the upper and lower waveguides have switched places,

\[
S_x \equiv \sqrt{1-\gamma_\text{SLR}} e^{i\phi_\text{SLR}} \sigma_x = \sqrt{1-\gamma_\text{SLR}} e^{i\phi_\text{SLR}} \begin{pmatrix}
0 & 1 \\
1 & 0
\end{pmatrix}
\]

All together we have the following matrix equation,

\begin{equation}
\boxed{
S_\text{SLR} = \begin{pmatrix}
S_{11} & S_{21} \\
S_{12} & S_{22}
\end{pmatrix}
= S_\text{MZI}^T \cdot S_x \cdot S_\text{MZI}
}
\label{eq:SLR}
\end{equation}

Let's consider the simple case where the reflector is made from a fixed lossless beamsplitter instead of an MZI. Substituting eqn~\ref{eq:bs} into~\ref{eq:SLR} we get,

\begin{equation}    
S_\text{SLR} = 
\begin{pmatrix}
2i \sqrt{\eta}\sqrt{1 - \eta} & (1- 2\eta) \\
(1- 2\eta) & 2i \sqrt{\eta}\sqrt{1 - \eta}
\end{pmatrix}
\sqrt{1-\gamma_\text{SLR}} e^{i\phi_\text{SLR}}
\label{eq:SLR2}
\end{equation}

Now we see that the reflection in $\frac{E_\text{r}}{E_\text{in}}$ in the main text is simply the $S_{11}$ component of the Sagnac loop reflector matrix above~\eqref{eq:SLR2}. Also, the Sagnac reflector behaves the same whether one inputs light in from port 1 or 2 (ie. $S_{11} = S_{22}, S_{12} = S_{22}$). Furthermore, we can see that the SLR is robust to beamsplitter design because the MZI matrix appears twice in the scattering matrix.

\subsection{Fabry–Pérot resonator}
\begin{figure}[hb!]
\centering\includegraphics[width=\linewidth]{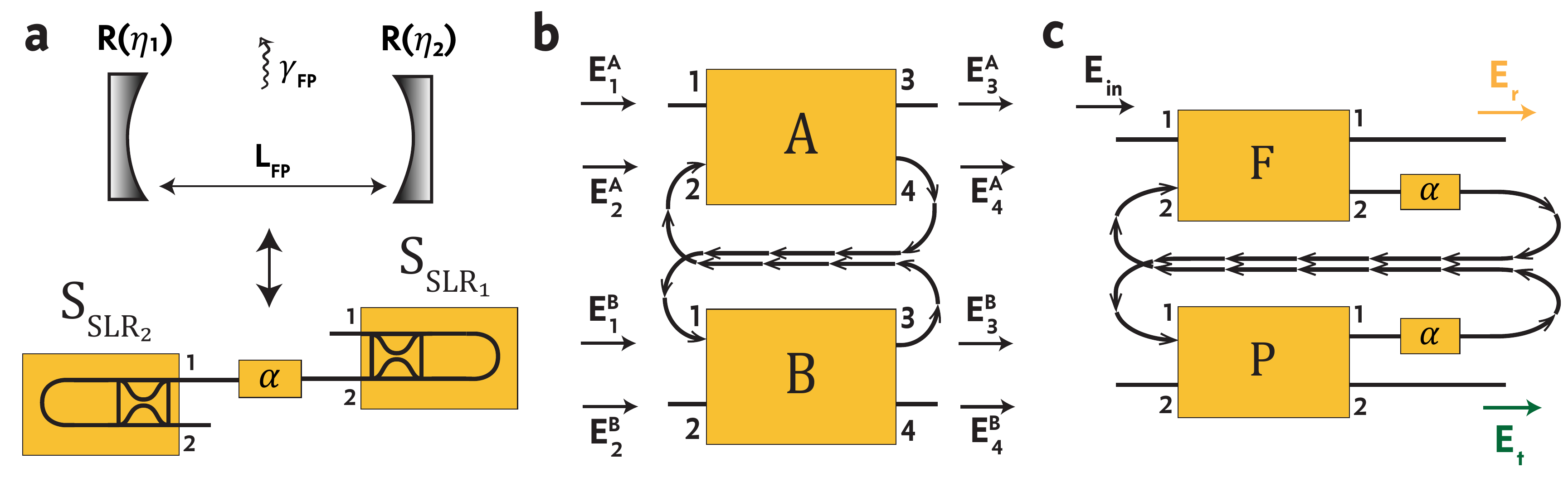} 
\caption{Fabry-Pérot modeling. (a) Equivalent pictures between mirrors and Sagnac loop reflectors. (b) Diagram of feedback between $A$ and $B$ matrices. (c) Forward propagating picture of Sagnac reflectors ($F$ and $P$). Feedback to form a resonance}
\label{fig:SI2}
\end{figure}

A resonance occurs when there is an internal feedback loop within a network of open scattering elements. Here, we connect one port of one Sagnac reflector (a single port serves as both input and output) to a different port of a second Sagnac reflector, illustrated in Figure~\ref{fig:SI2}a. This operation of connecting and then eliminating the internally connected ports can be described by a Redheffer star product, which is often used to describe electromagnetic field propagation in stratified, multilayered media~\cite{redheffer1959inequalities, liu2012s4, deltaRCWA}.

Before we explain the Redheffer star product, let's describe the Fabry-Pérot mode with a length of $L_\text{FP}$ and a single pass power loss of $\gamma_\text{FP}$. We define the single-pass amplitude loss and phase shift

\[
\alpha \equiv \sqrt{1-\gamma_\text{FP}} e^{i \phi_\text{FP}}
\]
where the phase shift is from wave propagation $\phi_\text{FP} = \beta(\omega) L_\text{FP}$. As seen in Figure~\ref{fig:SI2}a, this propagation term only shows up in one connected path of the Sagnac reflectors.

Now we can concisely describe the Fabry-Pérot matrix as follows,

\begin{equation}
\boxed{
S_\text{FP} = \left( 
\begin{pmatrix}
1 & 0 \\
0 & \alpha
\end{pmatrix} \cdot
 S_{\text{SLR}_1} \right) \star 
 \left( 
\begin{pmatrix}
\alpha & 0 \\
0 & 1
\end{pmatrix} \cdot
 S_{\text{SLR}_2} \right)
}
\label{eq:SFP}
\end{equation}
where $S_\text{SLR}$ are the matrices describing the MZI-controlled Sagnac loop reflectors in~\eqref{eq:SLR}. We explain this mysterious Redheffer star product $\star$ below.

\subsubsection{Redheffer star product}
Suppose we have two matrices as depicted in Figure~\ref{fig:SI2}b, $A$ and $B$ that each relate the incoming and outgoing waves:

\[
\begin{pmatrix}
E_3^{A} \\
E_4^{A}
\end{pmatrix}
=
A
\begin{pmatrix}
E_1^{A} \\
E_2^{A}
\end{pmatrix}
\quad \text{and} \quad
\begin{pmatrix}
E_3^{B} \\
E_4^{B}
\end{pmatrix}
=
B
\begin{pmatrix}
E_1^{B} \\
E_2^{B}
\end{pmatrix}
\]

We want to connect the lower waveguide of $A$ to the upper waveguide of $B$, which is to impose the following boundary condition,

\begin{equation}
E_4^{A} = E_1^{B}, \quad E_3^{B} = E_2^{A}
\label{eq:connect}
\end{equation}

These are the internal ports and will be eliminated. The matrices \( A \) and \( B \) are written as:

\begin{equation}
A =
\begin{pmatrix}
A_{31} & A_{32} \\
A_{41} & A_{42}
\end{pmatrix},
\quad
B =
\begin{pmatrix}
B_{31} & B_{32} \\
B_{41} & B_{42}
\end{pmatrix}
\label{eq:AB}
\end{equation}

The combined scattering matrix after connecting port 2 of \( A \) to port 1 of \( B \) is can be solved by writing out the full system of equations and then eliminating the connected variables. The result follows,

\begin{equation}
S_{\text{total}} =
\begin{pmatrix}
A_{31} + A_{32} \left( I - A_{42} B_{31} \right)^{-1} B_{31} A_{41} & A_{32} \left( I - A_{42} B_{31} \right)^{-1} B_{32} \\
B_{41} \left( I - A_{42} B_{31} \right)^{-1} A_{41} & B_{42} + B_{41} \left( I -A_{42} B_{31} \right)^{-1} A_{42} B_{32}
\end{pmatrix}
\label{eq:red}
\end{equation}
which is known as the Redheffer star product \( S_{\text{total}} = A \star B \). This is a remarkably concise method to describe resonances, and because this operation is associative ($A \star (B\star C) = (A \star B) \star C$), this allows for solving many cascaded Fabry-Perót cavities. 

\subsubsection{Detailed expansion}
To facilitate understanding, we choose to explicitly write out the terms in~\eqref{eq:SFP}.
Let's call the first SLR matrix $\bf F$ and the second SLR matrix $\bf P$. The $\alpha$ propagation term occurs when light travels from port 2 of $F$ to port 1 of $P$ and vice versa (Fig~\ref{fig:SI2}c). We internally connect these ports, which means our $A$ and $B$ matrices defined in~\eqref{eq:AB} now become,

\begin{equation}
A =
\begin{pmatrix}
F_{11} & F_{21} \\
\alpha F_{12} & \alpha F_{22}
\end{pmatrix},
\quad
B =
\begin{pmatrix}
\alpha P_{11} & \alpha P_{21} \\
P_{12} & P_{22}
\end{pmatrix}
\label{eq:ABFP}
\end{equation}
where the $F$ and $P$ terms are the $S_\text{SLR}$ terms written in~\eqref{eq:SLR}.

Finally we can substitute~\eqref{eq:ABFP} into~\eqref{eq:red} to get,
\[
S_{\text{FP}} =
\begin{pmatrix}
F_{11} + \frac{\alpha^2 F_{12} P_{11} F_{21}}{1 - \alpha^2 P_{11} F_{22}}
&
\frac{\alpha F_{12} P_{12}}{1 - \alpha^2 P_{11} F_{22}} 

 \\[10pt]

\frac{F_{21} P_{21} \alpha}{1 -\alpha^2 P_{11} F_{22}} 
&
P_{22} + \frac{\alpha^2 P_{21} F_{22} P_{12}}{1 - \alpha^2 P_{11} F_{22}}
\end{pmatrix}
\]

We see now how the terms take on the airy function form that arises from Fabry-Perót resonances. 

\section{Input-output model}
A more general approach of modeling a resonator that applies to any open quantum system is to start with the Heisenberg-Langevin equations. An oscillator may have an internal loss rate of $\kappa_i$, and a coupling rate to two baths $\kappa_{e1}$ and $\kappa_{e2}$. The equation of motion is written

\begin{equation}
\dot{a} = - \left(i \Delta + \frac{\kappa_\text{tot}}{2} \right)a - \sqrt{\kappa_{e1}}\alpha_\text{in} + \sqrt{\kappa_{e2}}\beta_\text{in}
\label{eq:inout}
\end{equation}
where $\kappa_\text{tot} = \kappa_i + \kappa_{e1}+\kappa_{e2}$, $\Delta$ is the frequency detuning from resonance, $\alpha$ and $\beta$ are the input photon flux from the first and second baths respectively.

In our measurements, we only probe the cavity through one path, $\alpha$. This gives us a boundary condition equations,

\begin{equation}
\alpha_\text{out} = \alpha_\text{in} + \sqrt{\kappa_{e1}}a,
\qquad 
\beta_\text{in} = 0
\label{eq:boundary}
\end{equation}

Since our system is quite symmetric, and we apply the same voltage to control each reflector of the Fabry-Perót cavity we can claim that the two external coupling rates are the same $\kappa_{e1} = \kappa_{e2} \equiv \kappa_e$. This allows us to solve for the steady state signal from our cavity when $\dot{a} = 0$,

\begin{equation}
\frac{\alpha_\text{out}}{\alpha_\text{in}} = 1 - \frac{\kappa_e}{i\Delta + \frac{\kappa_i + 2\kappa_{e}}{2}} = \frac{2i\Delta + \kappa_i}{2i\Delta + \kappa_i + 2\kappa_{e}}
\end{equation}

\section{MZIs with unbalanced beamsplitters}
In order to understand the MZI sensitivity to coupler fabrication, we simulate the maximum bar and cross outputs achievable with different beamsplitter ratios. We use the same $S_\text{MZI}$ matrix described previously as our model. 

\begin{figure}[ht!]
\centering\includegraphics[width=0.7\linewidth]{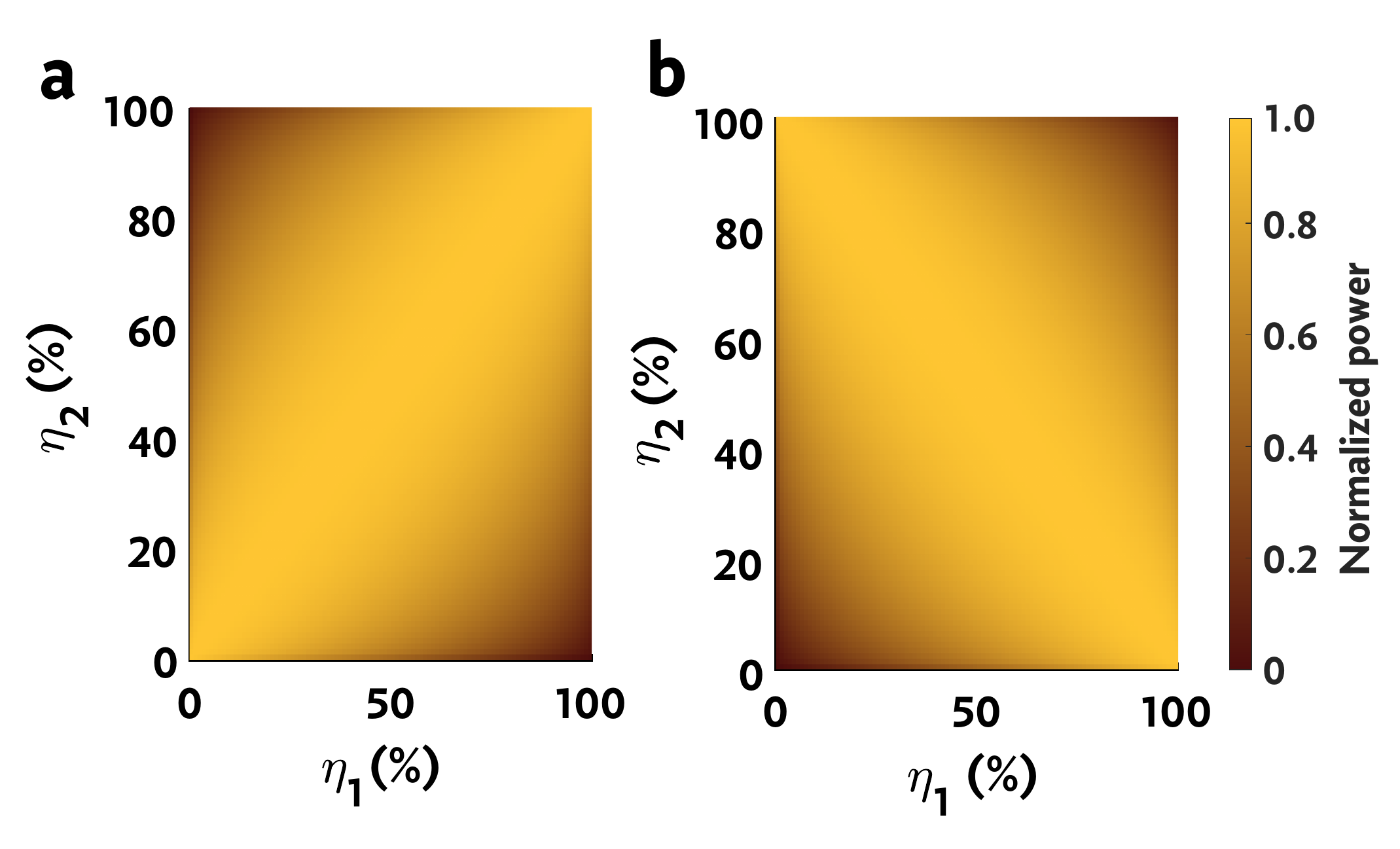} 
\caption{MZI outputs with any beamsplitter ratios. (a) Maximum achievable bar output of the MZI. (b) Maximum achievable cross output of the MZI}
\label{fig:SI3}
\end{figure}

Here, in Figure~\ref{fig:SI3}, we see that full power in the bar output requires equal couplers, while full power in the cross output (maximal extinction) requires couplers that add to unity. 

\section{More experimental details}
We can see a dark field image of the measured device, as well as a photograph of the measurement setup. The full spectrum report in the main text is shown here, in order to show the grating coupler response.

\begin{figure}[ht!]
\centering\includegraphics[width=\linewidth]{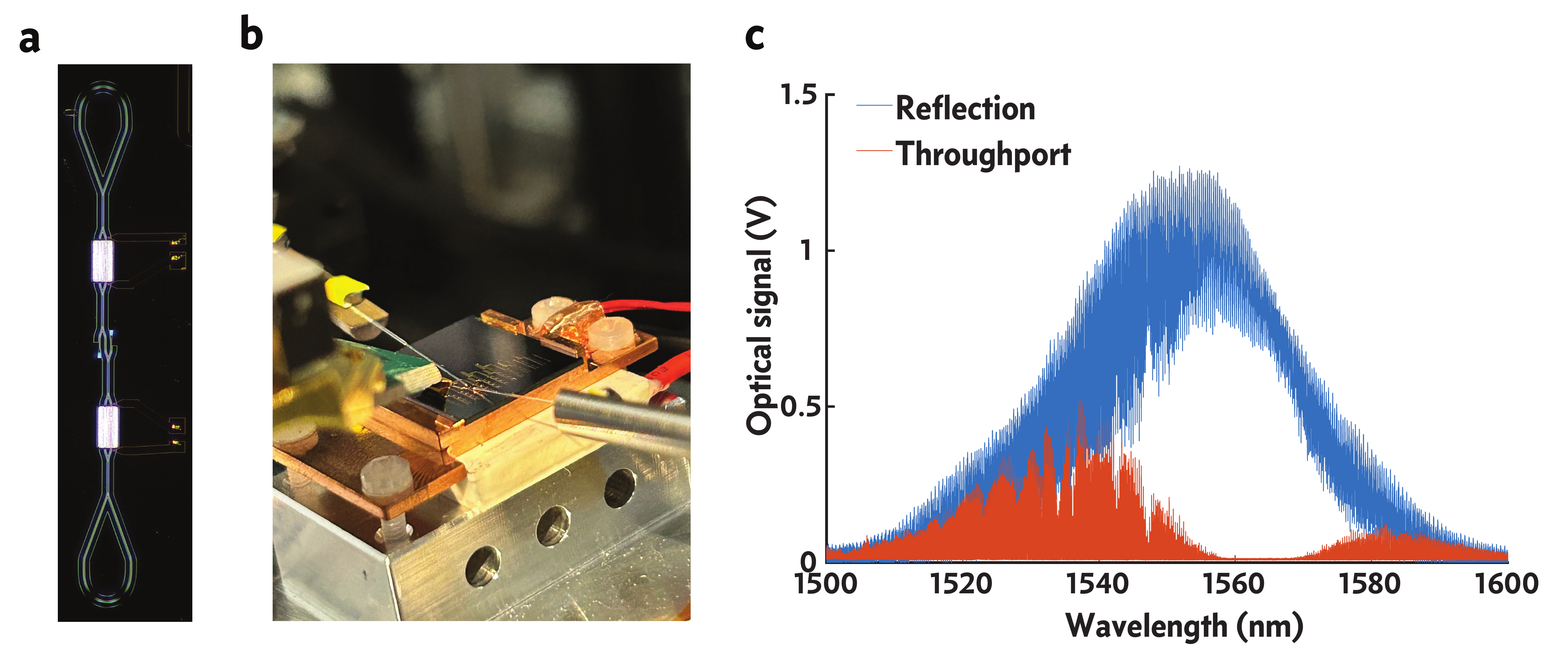} 
\caption{Device photos and data. (a) EvoCam image of MZI-controlled SLR Fabry-Perót cavity. (b) iPhone photograph of measurement setup. (c) Full spectrum of the device.}
\label{fig:SI4}
\end{figure}


\end{document}